\documentclass[11pt,reqno]{article}
\usepackage{amsmath}
\usepackage{amsfonts}
\usepackage{latexsym}
\usepackage[dvips]{graphicx}
\usepackage{epsf}

\allowdisplaybreaks \numberwithin{equation}{section}

\newcommand{\be}{\begin{equation}}
\newcommand{\ee}{\end{equation}}
\newcommand{\bea}{\begin{eqnarray}}
\newcommand{\eea}{\end{eqnarray}}

\newcommand{\Zom}{\mathbb{Z}}

\newcommand{\cC}{\mathcal{C}}

\newcommand{\cL}{\mathcal{L}}
\newcommand{\cM}{\mathcal{M}}

\newcommand{\SL}{\mathrm{SL}}
\newcommand{\SU}{\mathrm{SU}}

\newcommand{\vb}{\bar{v}}

\newcommand{\I}{\mathrm{i}}
\newcommand{\e}{\mathrm{e}}

\newcommand{\rmd}{{\rm d}}

\newcommand{\zb}{\bar{z}}

\newcommand{\vr}[1]{{\vec{r}\,}^{#1}}

\DeclareMathOperator{\im}{\mathrm{Im}}

\begin{document}

\begin{flushright} \small
ITP--UU--07/32 \\ SPIN--07/22 
\end{flushright}
\bigskip

\begin{center}
 {\LARGE\bfseries Membrane instantons from mirror symmetry }
\\[10mm]
Daniel Robles-Llana$^1$, Frank Saueressig$^1$, Ulrich Theis$^2$
and
Stefan Vandoren$^1$ \\[3mm]
$^1${\small\slshape
Institute for Theoretical Physics \emph{and} Spinoza Institute \\
Utrecht University, 3508 TD Utrecht, The Netherlands \\
{\upshape\ttfamily D.RoblesLlana, F.S.Saueressig,
S.Vandoren@phys.uu.nl} }
\\[3mm]
$^2${\small\slshape
Institute for Theoretical Physics \\
Friedrich-Schiller-University Jena, D--07743 Jena, Germany \\
{\upshape\ttfamily Ulrich.Theis@uni-jena.de} }\\

\end{center}
\vspace{5mm}

\hrule\bigskip

\centerline{\bfseries Abstract} \medskip

We use mirror symmetry to determine and sum up a class of membrane
instanton corrections to the hypermultiplet moduli space metric arising
in Calabi--Yau threefold compactifications of type IIA strings. These
corrections are mirror to the D1 and D($-1$)--brane instantons on the
IIB side and are given explicitly in terms of a single function in
projective superspace. The corresponding four-dimensional effective
action is completely fixed by the Euler number and the genus zero
Gopakumar--Vafa invariants of the mirror Calabi--Yau.

\bigskip

\hrule\bigskip

\section{Introduction}

Over recent years it has become clear that nonperturbative aspects
of string theory play an important role in low-energy effective
actions (LEEA) that are relevant for model building. An active area of
research has been to determine instanton corrections to
superpotentials in $N=1$ flux compactifications, since they can
cause moduli stabilization in vacua with a positive cosmological constant.
The computation of nonperturbative
superpotentials in string or M-theory was initiated in
\cite{Witten:1996bn,Harvey:1999as}, following ideas that were
developed in \cite{Becker:1995kb} for $N=2$ string
compactifications. In these approaches, one typically considers the first 
instanton correction that can be
computed using semiclassical methods in string theory. In general,
it is very hard to find expressions that are exact to all orders in
the string coupling constant $g_s$, unless one can make use of string 
dualities. For superpotentials in certain $N=1$ 
flux compactifications this was demonstrated
in \cite{Berglund:2005dm}, but for K\"ahler potentials this is more complicated
due to the absence of non-renormalization theorems (see however 
\cite{Grimm:2007xm} for some partial results).

In this paper we study nonperturbative phenomena in string theory in four 
dimensions with eight supercharges \cite{Becker:1995kb} and show that one can find
an exact solution for a class of instanton corrections to
the LEEA. That is, we are able to determine
and sum up an infinite series of membrane instanton
corrections to all orders in $g_s$. To obtain these exact results we make
use of various dualities
and symmetries in $N=2$ compactifications of string theories.
Moreover, in contrast to the K\"ahler potential in $N=1$ theories,
$N=2$ supersymmetry gives us full control over the perturbative
string loop corrections thanks to a non-renormalization theorem that
prevents corrections beyond one-loop \cite{RSV}.

The models we consider are type IIA strings
compactified on Calabi--Yau threefolds (CY) and their mirror
versions, type IIB compactified on the mirror CY, without any
fluxes turned on. The resulting LEEA
contains vector multiplets and hypermultiplets coupled to
(ungauged) $N=2$ supergravity. The vector multiplet moduli space is 
well understood in terms of 
mirror symmetry, and is not affected by inclusion of D-branes. 
Mirror symmetry between the hypermultiplet moduli spaces in IIA
and IIB is complicated by the fact that both moduli spaces receive
stringy perturbative and nonperturbative corrections. The
perturbative corrections are one-loop (as higher order corrections
can be absorbed in field redefinitions in the LEEA) and are proportional to the Euler number, consistent with
mirror symmetry \cite{RSV}. The nonperturbative corrections are
much more complicated, and their full form is unknown at present.
However, their generic structure was argued in
\cite{Becker:1995kb} to be as follows. On the type IIB side
nonperturbative corrections are given by instantons arising from
odd dimensional Euclidean D-branes wrapping even supersymmetric cycles,
together with NS5/D5-branes wrapping the entire CY. In type IIA
they are given by membranes wrapping supersymmetric three-cyles
and NS5-branes again wrapping the entire CY\@. We will consider
membrane instantons only, i.e., Euclidean D2-branes wrapping
supersymmetric three-cycles in the CY,
leaving the NS5-brane instantons for future research. We comment on 
how to determine these corrections in Figure 1. More
precisely, we consider membrane instantons coming
from D2-branes wrapping only half of the three-cycles, say the
"electric" or $A$-cycles. These are related to our recent results
on D1 and D($-1$) instanton corrections on the type IIB side
\cite{Robles-Llana:2006is} by mirror symmetry in the presence of D-branes
\cite{Ooguri:1996ck}. Geometrically, under the mirror map holomorphic 
two-cycles in the CY are exchanged by special Lagrangian three-cycles 
in the mirror CY. 

\begin{figure}[t!]
\setlength{\unitlength}{1cm}
\begin{center}
\begin{picture}(16,10)
\thicklines \put(3.5,9.4){\makebox(2,0.6)[c]{Hypermultiplet sector
$\cM_{\rm HM}$}} \put(12.5,9.4){\makebox(2,0.6)[c]{Vector
multiplet sector $\cM_{\rm VM}$}}
\put(0.5,8.2){\makebox(2,0.6)[c]{IIA / $X$}}
\put(7,8.2){\makebox(2,0.6)[c]{IIB / $Y$}}
\put(10.75,8.2){\makebox(2,0.6)[c]{IIA / $Y$}}
\put(14.25,8.2){\makebox(2,0.6)[c]{IIB / $X$}}
\put(10.75,0.2){\makebox(2,0.6)[c]{\{$\alpha^\prime$\}}}
\put(14.25,0.2){\makebox(2,0.6)[c]{\{$-$\}}}
\put(14.4,0.5){\vector(-1,0){1.7}}
\put(13.125,0.5){\makebox(1,0.6)[c]{mirror}}
\put(11.75,1.3){\line(0,-1){0.4}}
\put(11.75,1.3){\vector(-1,0){3.8}}
\put(9.5,1.3){\makebox(2,0.6)[c]{c-map}}
\put(5.4,1){\makebox(2,0.6)[l]{\{$\alpha^\prime$, $1\ell$\}}}
\put(6,1.7){\vector(0,1){1}}
\put(6,1.8){\makebox(2,0.6)[c]{SL(2,$\Zom$)}}
\put(0.13,2.8){\makebox(2,0.6)[l]{\{$1\ell$, A-D2\}}}
\put(5.4,2.8){\makebox(2,0.6)[l]{\{$\alpha^\prime$, D($-1$),
D1\}}}
 \put(0.75,3.5){\vector(0,1){1.1}}
\put(1,3.7){\makebox(2,0.6)[l]{e/m duality}}
\put(0,4.8){\makebox(2,0.6)[l]{ \{$1\ell$, A-D2, B-D2\} }}
\put(5.4,4.8){\makebox(2,0.6)[l]{\{$\alpha^\prime$, D($-1$), D1,
D3, D5\} }}
 \put(6,5.5){\vector(0,1){1.1}}
\put(6,5.7){\makebox(2,0.6)[c]{SL(2,$\Zom$)}}
\put(5.4,6.8){\makebox(2,0.6)[l]{\{$\alpha^\prime$, D($-1$), D1,
D3, D5, NS5\} }} \put(0,6.8){\makebox(2,0.6)[l]{ \{$1\ell$, D2,
NS5\} }}
\put(5.1,3.1){\vector(-1,0){1.7}}
\put(3.75,3.1){\makebox(1,0.6)[c]{mirror}}
\put(3.4,5.1){\vector(1,0){1.7}}
\put(3.75,5.1){\makebox(1,0.6)[c]{mirror}}
\put(5.1,7.1){\vector(-1,0){1.7}}
\put(3.75,7.1){\makebox(1,0.6)[c]{mirror}}
\put(0,8){\line(1,0){16}} \put(10.7,0){\line(0,1){1}}
\put(10.7,2){\line(0,1){8}}
\end{picture}
\end{center}
\parbox[c]{\textwidth}{\caption{\label{zwei}{\footnotesize
Prospective duality chain for determining the full quantum LEEA of type II
strings compactified on a generic CY $X$, and its mirror partner $Y$. 
In the vector multiplet
sector there are $\alpha'$ corrections that appear on the IIA side
only and can be obtained via mirror symmetry; they comprise  worldsheet loop
and instanton corrections. The c-map transfers these
into the IIB hypermultiplet sector. In addition, there is a
one-loop $g_s$ correction, in the figure denoted by $1\ell$, determined
in \cite{RSV}. 
Imposing $\SL(2,\Zom)$ invariance
produces the nonperturbative corrections arising from D1-brane and
more general $(p,q)$-string instantons as well as D($-1$)
instantons \cite{Robles-Llana:2006is}. 
The latter naturally combine with the perturbative
$\alpha'$ and $g_s$ corrections. As shown in this paper, applying
mirror symmetry to these corrections gives rise to the $A$-cycle
D2-brane instanton contributions on the IIA side. Though beyond the scope
of this paper, one may now
continue to employ various dualities that should in principle
produce all possible quantum corrections: using electromagnetic
(e/m) duality to impose symplectic invariance will give the $B$-cycle
D2-brane instantons. Mirror symmetry will map these to the as of
yet unknown D3- and D5-brane instanton corrections on the IIB
side. Another application of $\SL(2,\Zom)$ duality then will give
rise to pure NS5-brane and D5--NS5 bound state instantons.
Finally, applying mirror symmetry one last time will produce the
NS5-brane corrections on the IIA side.}}}
\end{figure}
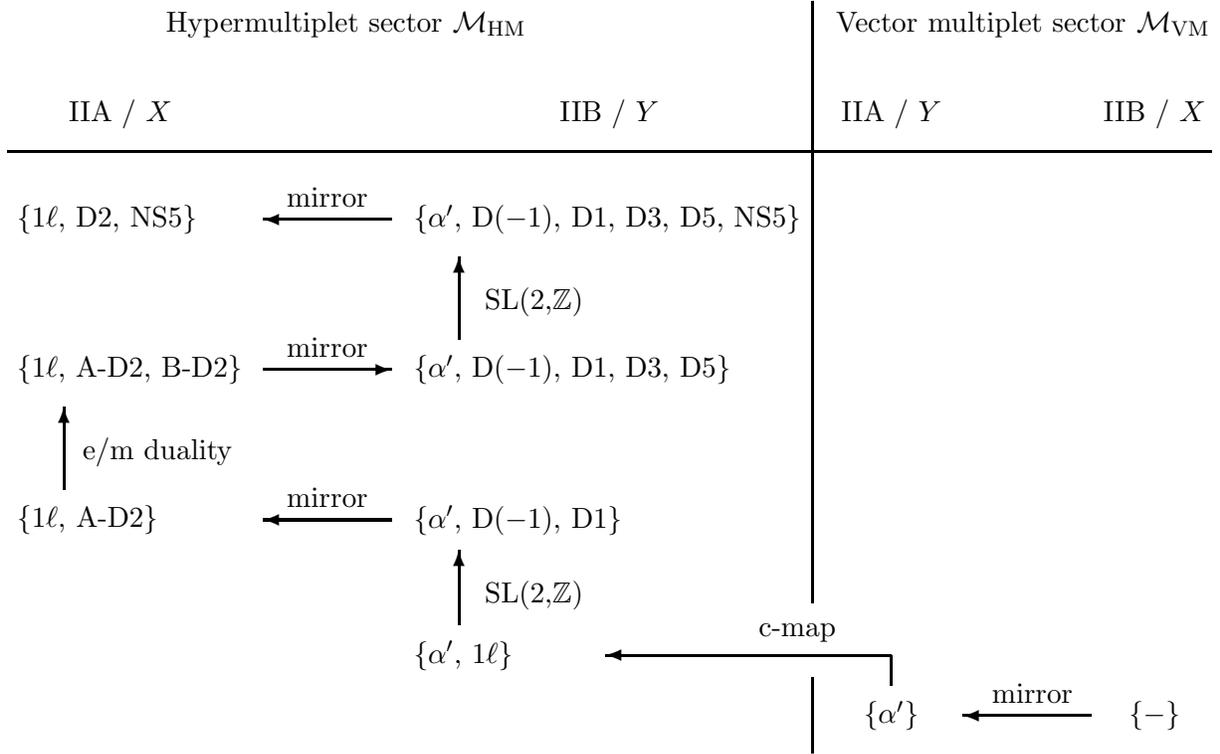

The main result of this paper is the implementation of the mirror map
and the explicit form of the instanton corrected hypermultiplet geometry.
In fact, recently the instanton 
corrections were determined in \cite{SV} in the conifold limit of type IIB
strings upon shrinking the size of a two-cycle. The results showed
precise agreement with the membrane instantons obtained in
\cite{Ooguri:1996me} in type IIA strings on the mirror CY, thereby
providing a nonperturbative test of mirror symmetry in the conifold
limit. Here, we assume mirror symmetry to hold at a generic point
in the moduli space, and use it to determine the instanton corrections
to the  effective action away from the conifold limit, where the
hypermultiplet  moduli space is quaternion-K\"ahler instead of
hyperk\"ahler.

Finding the nonperturbative corrections to the hypermultiplet
moduli space by means of a microscopic instanton calculation has
been a notoriously complicated problem. However, for rigid CY manifolds 
($h_{1,2}=0$), yielding only the universal hypermultiplet, a microscopic
string theory calculation was performed in \cite{Davidse:2005ef}
following the procedure outlined in \cite{Becker:1995kb}. In this
approach, the one-instanton sector can be determined in the semiclassical
approximation, up to an overall numerical coefficient that comes out of
the one-loop determinant of fluctuations around the instanton. The
method used in the present paper fixes this coefficient in terms of the Euler number of the CY and allows us to
go beyond the semiclassical approximation.

Finally, we wish to point out that if the CY is $K3$-fibered, our
results are related to a class of worldsheet instantons in the heterotic
string on $K3\times T^2$ using string duality. The mapping of membrane
instantons to worldsheet instantons was recently investigated in
\cite{Halmagyi:2007wi}, and it would be interesting to study further how
our results fit into that framework.

In the next section, we briefly review some general aspects of off-shell
effective actions for hypermultiplet and tensor multiplet couplings to
$N=2$ supergravity that are needed in later sections. In Section 3 we
present the connection with the string theory variables for both the
type IIA and IIB theories and discuss the breaking of isometries by
instantons. Section 4 deals with microscopic aspects of mirror symmetry, 
including both the NS-NS and R-R sectors, and we construct the mirror
map between the IIA and IIB hypermultiplet moduli spaces. Finally, in
Section 5, we implement mirror symmetry to determine the membrane
instanton correction starting from previously known results on D1 and
D($-1$) instantons in IIB. We close with some conclusions and further
discussions on Section 6. Two appendices are included at the end with technical details and useful 
formulas.

\section{Off-shell effective actions}
\label{sec:superspace}
%
\begin{figure*}[t]
\setlength{\unitlength}{1cm}
\begin{center}
\begin{picture}(12,3.3)
\thicklines
\put(2,3){\makebox(2,0.6)[c]{IIA / $X$}}
\put(9,3){\makebox(2,0.6)[c]{IIB / $Y$}}
\put(3,3){\line(0,-1){0.4}} \put(10,3){\line(0,-1){0.4}}
\put(1.25,2.6){\line(1,0){3.5}} \put(8.25,2.6){\line(1,0){3.5}}
\put(1.25,2.6){\vector(0,-1){0.5}}
\put(4.75,2.6){\vector(0,-1){0.5}}
\put(8.25,2.6){\vector(0,-1){0.5}}
\put(11.75,2.6){\vector(0,-1){0.5}}
\put(0.25,1.4){\makebox(2,0.6)[c]{$\mathcal{M}_{\rm VM}^A$}}
\put(3.75,1.4){\makebox(2,0.6)[c]{$\mathcal{M}_{\rm HM}^A$}}
\put(7.25,1.4){\makebox(2,0.6)[c]{$\mathcal{M}_{\rm VM}^B$}}
\put(10.75,1.4){\makebox(2,0.6)[c]{$\mathcal{M}_{\rm HM}^B$}}
%
\put(0.25,0.7){\makebox(2,0.6)[c]{$2 h_{1,1}(X)$}}
\put(3.75,0.7){\makebox(2,0.6)[c]{$4 (h_{1,2}(X) + 1)$}}
\put(7.25,0.7){\makebox(2,0.6)[c]{$2 h_{1,2}(Y)$}}
\put(10.75,0.7){\makebox(2,0.6)[c]{$4 (h_{1,1}(Y) + 1)$}}
%
\put(0.25,0){\makebox(2,0.6){$\alpha^\prime$}}
\put(3.75,0){\makebox(2,0.6){$g_s$}}
\put(7.25,0){\makebox(2,0.6){$-$}}
\put(10.75,0){\makebox(2,0.6){$\alpha^\prime, g_s$}}
%
%
%
\end{picture}
\end{center}
\parbox[c]{\textwidth}{\caption{\label{eins}{\footnotesize
Massless matter spectrum arising in CY compactifications of
type II strings. The moduli spaces for vector and hypermultiplets
are denoted by ${\cal M}_{\rm VM}$ and ${\cal M}_{\rm HM}$
respectively, with their real dimensions below in terms of the
Hodge numbers $h_{p,q}$ of the CY. The last line indicates
possible quantum corrections to the respective sector. Here
$\alpha^\prime$ and $g_s$ denote corrections from the worldsheet
conformal field theory and in the (four-dimensional) string coupling
constant, respectively. Note that the vector multiplet sector of the
type IIB compactification is classically exact.}}}
\end{figure*}
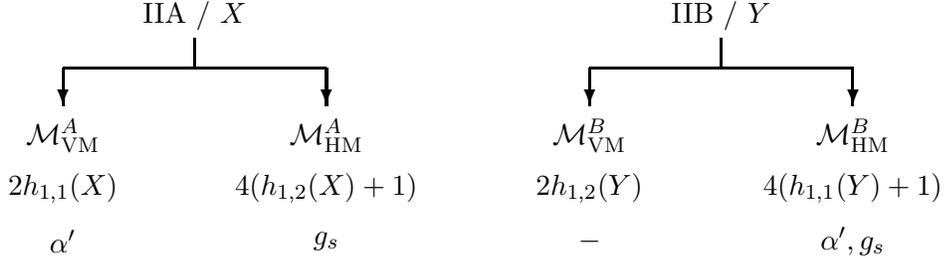

In this section, we describe general aspects of superspace LEEA for
hypermultiplets and their dual tensor multiplets. After a short review
of projective superspace, we give the resulting tree-level and one-loop
effective actions for type II strings compactified on a CY\@. 

The total moduli space ${\mathcal M}$ for a type II string
compactification on a generic CY locally factorizes into a direct product
${\mathcal M}={\mathcal M}_{\rm VM} \times {\mathcal M}_{\rm HM}$, where
${\mathcal M}_{\rm VM}$ and ${\mathcal M}_{\rm HM}$ denote the vector and
hypermultiplet moduli space. The metrics on ${\mathcal M}^{\rm IIA}_{\rm
VM}/X$ and ${\mathcal M}^{\rm IIB}_{\rm VM}/Y$ are entirely determined by
the prepotentials $F(X)$ on the moduli spaces of (complexified) K\"ahler 
deformations on $X$ and complex structure deformations on $Y$,
respectively. As the four-dimensional dilaton belongs to a hypermultiplet,
the generic factorization of the moduli space implies that the vector
moduli spaces are exact at string tree-level. This is summarized in 
Figure 2. 

The scalars of the hypermultiplets coupled to $N=2$ supergravity
parametrize a quaternion-K\"ahler manifold \cite{Bagger:1983tt}. 
For type IIA (type IIB) compactifications, this manifold is spanned by
$h_{1,2}+1$ ($h_{1,1}+1$) physical hypermultiplets. Generic hypermultiplet
couplings do not admit an off-shell description with a finite number of
auxiliary fields, so no simple superspace formula exists for the effective
action. As we explain in the next section, for the problem at hand, there
will be a suitable number of Peccei--Quinn-like isometries that act as
shifts on some of the scalars. In this case, the hypermultiplets have a
dual description in terms of $N=2$ tensor multiplets which allow for an
off-shell description with a finite number of auxiliary fields in
projective superspace \cite{Gates:1984nk,Karlhede:1984vr}. The $N=2$
tensor multiplets are written as
\begin{equation}
\eta^I(\zeta)=\frac{v^I}{\zeta}+x^I-{\bar v}^I\zeta\ ,
\end{equation}
where $v$ and $x$ denote $N=1$ chiral and real linear (containing a real
scalar and a tensor) superfields respectively, while $\zeta$ is a complex
coordinate on the Riemann sphere. In rigid $N=2$ supersymmetry, superspace
Lagrangian densities for the tensor multiplets can be written as
\begin{equation}\label{cint}
{\cal L}(x^I,v^I,{\bar v}^I)=\im\oint_{\mathcal C}\, \frac{{\rm d}\zeta}{2\pi
i\zeta}\, H \big(\eta^I(\zeta),\zeta\big)\ ,
\end{equation}
with ${\mathcal C}$ an appropriately chosen contour that typically
encloses the poles and branch cuts of the function $H$. Supersymmetry
does not pose any constraints on ${\mathcal C}$. By construction $\cL$, 
when taken as a function of the scalar fields, automatically satisfies the
constraint for rigid supersymmetry
\be\label{Llap}
\cL_{x^I x^J} + \cL_{v^I \vb^J} = 0\ , 
\ee
where ${\cal L}_{x^I}$ denotes the derivative of ${\cal L}$ with
respect to $x^I$, etc.

The tensor multiplet action is obtained by integrating the superspace
Lagrangian ${\cal L}$ over half of superspace, with coordinates
$\theta_{i\alpha},{\bar \theta}^i_{\dot\alpha}$, where $i=1,2$ is an
$\SU(2)_R$ index and $\alpha,\dot\beta=1,2$ are Lorentz spinor indices.
For vector multiplet actions, one integrates superspace densities over a
chiral subspace spanned by the $\theta_{i\alpha}$ only. Such terms are
called F-terms. For tensor multiplets, one chooses a different subspace
spanned by, say, $\theta_{1\alpha}$ and ${\bar \theta}^2_{\dot \alpha}$.
Such terms can be called twisted F-terms, and the action for the tensor
multiplets then takes the form
\begin{equation}
S=\int {\rm d}^4x\,{\rm d}^2{\theta}\,{\rm d}^2{\bar
\theta}\,\,{\cal L}\ .
\end{equation}
For heterotic superstrings on $K3\times T^2$, which are dual to
our models, projective superspace arises naturally from the hybrid
formalism \cite{Linch:2006sh}, whereas for type II strings on a CY
the situation is less clear; see \cite{Kappeli:2006fj} for more details.
Superspace effective actions have also been studied in
\cite{Berkovits:1995cb}.

To couple the resulting tensor multiplet action to supergravity,
we use the superconformal calculus and introduce a compensating
tensor multiplet. The theory can then be made scale invariant, and
the constraints from conformal symmetry lead to the conditions
that $H$ has no explicit $\zeta$ dependence and is homogeneous of
degree one under the contour integral \cite{deWit:2001dj}. The
resulting expression is then conformally coupled to the Weyl
multiplet, which contains the degrees of freedom of $N=2$
conformal supergravity; it is the gauge multiplet of the $N=2$
superconformal algebra. Gauge-fixing the $\SU(2)_R$ and
dilatations eventually gives the tensor multiplet couplings to
Poincar\'e supergravity in component language \cite{deWit:2006gn}.
What is important here is that the entire supergravity action is
determined by a single function ${\cal L}(x,v,\bar v)$ of the tensor
multiplet scalar components.

At the superconformal level the dualization of the tensor multiplets 
to hypermultiplets can be done by performing a Legendre transform on
the superspace Lagrangian density ${\cal L}$ \cite{Hitchin:1986ea}. The
resulting function is the hyperk\"ahler potential on the hyperk\"ahler
cone describing superconformally coupled hypermultiplets
\cite{deWit:1999fp}. Taking the superconformal quotient as in
\cite{deWit:2001dj} leads to the quaternion-K\"ahler hypermultiplet
moduli space.

Before doing the Legendre transform, one can construct a tensor potential
\cite{deWit:2006gn}
\begin{equation}\label{TP}
\chi(x^I,v^I,{\bar v}^I)=x^I{\cal L}_{x^I} - {\cal L}\ ,
\end{equation}
which satisfies the differential relations
\begin{equation} \label{cLrel}
 \cL_{x^I x^J} = \tfrac{1}{2} \left( \chi_{x^I x^J} +
\chi_{v^I \vb^J} \right) \, .
\end{equation}
The two functions $\chi$ and ${\cal L}$ will play the central role in our
discussion; they encode all the quantum corrections, from string loops as
well as from instantons. It turns out to be particularly convenient to work
with $\chi$, since symmetries of the effective action directly translate to
invariances of the tensor potential. In Section \ref{sec:5}, we will give
explicit formulas for the instanton corrections to both of these functions.
Further details will also be given in Appendix B.
\begin{table}[t]
\begin{center}
\begin{tabular}{c|c}
\hspace*{1cm} type IIA strings on $X$ \hspace*{1cm} & \hspace*{1cm} type
IIB strings on $Y$ \hspace*{1cm} \\[1.1ex] \hline
$I  = 0,1,\dots,h_{1,2}(X)+1$ & $I  = 0,1,\dots,h_{1,1}(Y)+1$ \\[1.1ex]
$\Lambda = 1,\dots,h_{1,2}(X)+1$ & $\Lambda = 1,\dots,h_{1,1}(Y)+1$ \\[1.1ex]
$a = 2,\dots,h_{1,2}(X)+1$ & $a = 2,\dots,h_{1,1}(Y)+1$
\end{tabular}
\end{center}
\parbox[c]{\textwidth}{\caption{\label{t.1}{\footnotesize
Summary of indices used in the hypermultiplet sector. The index
$a$ runs over one value less than $\Lambda$, making the
coordinates $z^a$ projective, and $I=\{0,\Lambda\}$ also includes
the compensator. For $X$ and $Y$ mirror to each other,
$h_{1,2}(X)=h_{1,1}(Y)$ so that the number of multiplets agrees.}}}
\end{table}

At string tree-level, the hypermultiplet moduli space can be determined
from the vector multiplet effective action in the T-dual theory by
applying the c-map \cite{Cecotti:1988qn}. The resulting quaternion-K\"ahler
geometry was worked out in \cite{Ferrara:1989ik}, and the corresponding
superspace Lagrangian takes the simple form \cite{Rocek:2005ij,Rocek:2006xb}
(also see \cite{Berkovits:1995cb})
\begin{equation}\label{classL}
{\cal L}=\im\oint_{\mathcal C_0} \frac{{\rm d}\zeta}{2\pi
\I\zeta}\frac{F(\eta^\Lambda)}{\eta^0}\ .
\end{equation}
Here, $F$ is the vector multiplet prepotential of the T-dual theory,
homogeneous of degree two, but with the vector multiplets $X^\Lambda$
replaced by tensor multiplets $\eta^\Lambda$. (See Table \ref{t.1} for
index ranges.) Finally, $\eta^0$ can be chosen to be the conformal
compensator. Equation \eqref{classL} was first derived in a certain
$\SU(2)_R$ gauge $(v^0=0)$ with the contour $\mathcal C_0$ chosen around
the origin. It was then rederived in \cite{Neitzke:2007ke} in a
gauge-independent way with the contour $\mathcal C_0$ around one of the
zeroes, say $\zeta_+$, of $\zeta\eta^0$. In the gauge $v^0=0$ one
recovers the results of \cite{Rocek:2005ij}.

At one-loop order in $g_s$, the resulting action is given by
\cite{RSV}\footnote{We use a slightly different
normalization for the prepotential than in \cite{RSV}. As
a result, the overall numerical coefficient in \eqref{oneloopL} relative
to the classical term \eqref{classL} is different. With the conventions
used in this paper, the monodromies of the prepotential around the conifold
point are integer-valued, as is common in the literature.}
\begin{equation}\label{oneloopL}
{\cal L}=\pm \im \frac{\I\chi_E}{24\pi}\oint_{\mathcal C_1} \frac{{\rm
d}\zeta}{2\pi \I\zeta}\,\,\eta^0\,{\rm ln}\,\eta^0\ ,
\end{equation}
with the positive sign for IIB and the negative sign for type IIA. Here,
$\chi_E$ is the Euler number of the CY\@. Notice that the results for
type IIA and IIB are consistent with mirror symmetry, since the Euler
number switches sign for the mirror CY\@. The contour $\mathcal C_1$ is
taken around the logarithmic branch cut between the origin and $\zeta_+$
\cite{Alexandrov:2007ec}. The formula \eqref{oneloopL} generalizes the
results for the universal hypermultiplet obtained in
\cite{Antoniadis:2003sw,Anguelova:2004sj}. Moreover, in
\cite{Antoniadis:2003sw} it was shown that higher loop corrections
can be absorbed by field redefinitions. The effective absence of
higher-loop corrections was given an explanation in
\cite{RSV} in terms of a nonrenormalization
theorem in projective superspace, where it can be generalized to an
arbitrary number of hypermultiplets.

\section{String theory variables}

In the previous section, we gave generic formulas for tensor-/hypermultiplet
LEEA in string perturbation theory. To connect these formulas to a specific
string theory compactification we need to specify the relation between the
tensor multiplet components and the four-dimensional physical fields arising
from the dimensional reduction of the ten-dimensional massless modes, which
are given by conformally and SU(2)$_R$ invariant combinations of the tensor
multiplet scalars. In this section, we state these relations for type IIB
and type IIA compactifications, and we analyze the breaking of isometries
due to instantons. In view of the discussion of mirror symmetry in the 
next section it is thereby useful to distinguish between the NS-NS and RR
sectors of the compactification.

\subsection{Type IIB}

In type IIB compactifications on a CY manifold $Y$ the four-dimensional
fields organize themselves into the gravitational multiplet, $h_{1,2}(Y)$
vector multiplets, $h_{1,1}(Y)$ tensor multiplets, and a double-tensor
multiplet \cite{BGHL}. The latter contains two tensors and two scalars
and has an off-shell description in terms of two tensor multiplets,
which contain the four compensating scalar fields that restore the
dilatation and SU(2)$_R$ symmetries. Thus, off-shell there are $h_{1,1}
(Y)+2$ tensor multiplets.

The NS-NS sector of the compactification contains the ten-dimensional
dilaton $\phi$ and the complexified K\"ahler moduli $z^a$ arising from
integrating the K\"ahler form $J$ and the NS two-form $ B_{\rm NS}$ over
a basis of two-cycles $\gamma^a_{(2)}$ of $H_2(Y,{\mathbb Z})$
\begin{equation}\label{kahlercoords} 
z^a=b^a + \I t^a = \int_{\gamma^a_{(2)}}\!\! \big( B_{{\rm NS}}+\I J
\big)\ ,\qquad a=2,...,h_{1,1}+1\ .
\end{equation}
The dilaton provides one of the scalars of the double tensor multiplet,
while the $z^a$ account for $2 h_{1,1}(Y)$ of the tensor multiplet scalars.

In the RR sector, the axion coming from the zero form $C_0$ corresponds to
the second scalar in the double-tensor multiplet, while the periods of the
RR two-form $C_2$,
\begin{equation}
c^a=\int_{\gamma^a_{(2)}}\! C_2\ ,
\end{equation}
make up the remaining $h_{1,1}(Y)$ scalars. Furthermore, the
four-dimensional tensor fields are provided by the space-time parts of
$B_{\rm NS}$ and $C_2$ together with $h_{1,1}(Y)$ tensors arising as
the periods of the RR four-form $C_4$ with respect to $\gamma_{(2)}^a$.
The dilaton-axion system can be combined into the complex scalar
\begin{equation}
\tau = \tau_1 + \I \tau_2 =C_0 + \I\mspace{2mu} \e^{-\phi}\ .
\end{equation}

The relation between the ``microscopic'' scalars $\tau$, $z^a$, $c^a$
and the scalars appearing in the tensor multiplets \eqref{classL} is
given by \cite{Neitzke:2007ke}
\begin{equation}\label{vocabIIB}
\tau = \frac{1}{(r^{0})^2} \left(\vr{0} \cdot \vr{1} + \I \, |
\vr{0} \times \vr{1} | \right)\ ,\qquad  z^a =
\frac{\eta_+^a}{\eta_+^1}\ ,\qquad c^a = \frac{(\vr{0} \times
\vr{1}) \cdot (\vr{1} \times \vr{a})}{|\vr{0} \times \vr{1}|^2} \ .
\end{equation}
Here we have introduced
\begin{equation}\label{SU2vec}
\vr{I} = \left[2 \, v^I, \, 2 \, \vb^I, \, x^I \right] \, , \qquad
\vr{I}\! \cdot \vr{J} = 2 v^I \vb^J + 2 v^J \vb^I + x^I x^J\ ,
\end{equation}
and $\eta^\Lambda_+=\eta^\Lambda(\zeta_+)$, where $\zeta_+$ is one
of the roots of $\zeta\eta^0(\zeta)$. See Appendix \ref{App:B} for more
details. 

In the large volume limit the $b^a$ and $c^a$ are subject to
continuous Peccei--Quinn shift symmetries. Nonperturbatively, these
symmetries are broken to discrete subgroups by worldsheet instantons
and D1 instantons, respectively. Similarly, the shift symmetry of the
axion $\tau_1$ is broken by D($-1$) instantons. In our set-up, we are
still left with $h_{1,1}+2$ tensors. Dualizing these tensors to scalars
again leads to $h_{1,1}+2$ continuous shift symmetries. These will also
be broken by instantons; the first $h_{1,1}$ by D3-brane instantons, and
the two remaining ones, originating from the double-tensor multiplet,
by D5 and NS5-brane instantons. Including these instanton corrections  
requires a formulation in terms of hypermultiplets which are not dual to
tensor multiplets, so the tensor multiplet framework used in this paper
is no longer applicable. 

\subsection{Type IIA}

When compactifying type IIA strings on $X$ the off-shell formulation of
the hypermultiplet sector uses $h_{1,2}(X)+2$ tensor multiplets. Not
counting the conformal compensators, consisting of four scalars, we
therefore have $3h_{1,2}+2$ physical scalars and $h_{1,2}+2$ tensors.

The NS-NS sector contains the dilaton $\phi$ and the complex structure
deformations of $X$, which together account for $2h_{1,2}(X)+1$ scalars.
In order to find the coordinates on the complex structure moduli space,
we introduce a symplectic basis of $h_{1,2}(X)+1$ $A$- and $B$-cycles 
$\{\gamma^\Lambda_{(3)},\gamma_\Lambda^{(3)}\}$ of $H_3(X,{\mathbb Z})$. 
The complex structure deformations can then be parametrized by the
periods of the holomorphic three-form $\Omega$ with respect to the
$A$-cycles $\gamma^\Lambda_{(3)}$. Choosing $\gamma^1_{(3)}$ to be the
cycle dual to $\Omega$, the inhomogeneous (physical) complex structure
moduli $z^a$ are defined as
\begin{equation}\label{specialcoords}
X^\Lambda = \int_{\gamma^\Lambda_{(3)}}\! \Omega\ ,\qquad
z^\Lambda\equiv\frac{X^\Lambda}{X^1}=\big(1,z^a\big)\ .
\end{equation}

In the RR sector, the periods of the RR three-form $C_3$
\begin{equation}
A^\Lambda =\int_{\gamma^\Lambda_{(3)}}\! C_3\ ,
\end{equation}
give the remaining $h_{1,2}(X)+1$ physical scalars. In principle,
integrating $C_3$ over the dual $B$-cycles gives rise to $h_{1,2}(X)+1$
additional scalars $B_\Lambda$. In our framework these are dualized into
tensors. Together with the space-time part of $B_{\rm NS}$ this gives
$h_{1,2}+2$ tensors, so the total number of degrees of freedom works
out correctly. Notice that upon interchanging $h_{1,1}$ with $h_{1,2}$,
the IIA spectrum precisely agrees with the one for the type IIB theory.
This is of course a manifestation of mirror symmetry. At the classical 
level, the mirror map relating the tree-level IIA and IIB hypermultiplet 
moduli spaces was found to be \cite{BGHL} 
\begin{equation}\label{Cl-MM}
\phi_{{\rm IIA}}=\phi_{{\rm IIB}}\ ,\qquad A^1 = \tau_1\ ,\qquad A^a =
-(c^a - \tau_1 b^a)\ ,\qquad z^a_{{\rm IIA}} = z^a_{{\rm IIB}}\ .
\end{equation}
At the classical level, the RR fields $A^\Lambda$ have continuous Peccei-Quinn isometries which will be broken to a discrete subgroup by the instanton corrections. Using \eqref{vocabIIB}, the type IIA scalars $\e^{-\phi}$, $z^a$,
$A^\Lambda$ are related to the tensor multiplet components via
\be\label{vocabIIA}
\e^{-\phi} = \frac{1}{(r^0)^2}\, |\vr{0} \times \vr{1}|\ , \qquad z^a =
\frac{\eta^a_+}{\eta^1_+}\ ,\qquad A^\Lambda = \frac{1}{(r^0)^2}\,
(\vr{0} \cdot\vr{\Lambda})\ .
\ee
In the next section, we show that the mirror map \eqref{Cl-MM} in fact
also holds in the presence of instantons, but with additional periodic identifications of the fields.

Dualizing the scalars $B_\Lambda$ into tensor fields and keeping the NS
two-form makes the associated shift symmetries manifest. Nonperturbatively
these symmetries are broken by membrane instantons that wrap the
$B$-cycles and by NS5-brane instantons, respectively. These instanton
corrections are not taken into account in our analysis, since their
inclusion cannot be described by our off-shell tensor multiplet description.

\section{Mirror symmetry}
\label{sec:mirror}

Mirror symmetry can be phrased as the statement that the A-model topological
amplitudes on a CY $X$ can be mapped to the B-model topological amplitudes
of the mirror CY $Y$. At the level of the four-dimensional LEEA this implies
that the vector multiplet moduli spaces ${\mathcal M}^{\rm IIA}_{\rm VM}/Y$
and ${\mathcal M}^{\rm IIB}_{\rm VM}/X$ are isomorphic, i.e., there is a
choice of holomorphic coordinates for which the prepotentials (computed
from the genus zero topological amplitude) underlying these spaces agree. 
Via the c-map \eqref{classL} this entails that the hypermultiplet moduli
spaces ${\mathcal M}^{\rm IIA}_{\rm HM}/X$ and ${\mathcal M}^{\rm
IIB}_{\rm HM}/Y$ agree at \emph{string tree-level}. This version of mirror
symmetry is by now well established, see e.g.\ \cite{mirrorbook}.

For the purpose of this paper in which we also include D-branes, we need a
version of mirror symmetry that includes D-branes, in particular wrapped
Euclidean D-branes representing spacetime instantons. At the level of the
LEEA of type IIA/IIB string theory, this version of mirror symmetry states that 
${\mathcal M}^{\rm IIA}_{\rm HM}/X$ and  ${\mathcal M}^{\rm IIB}_{\rm
HM}/Y$ are also isomorphic once nonperturbative corrections are included. 
In this section, we assume mirror symmetry to hold and determine the
mirror map to relate the IIB and IIA hypermultiplet moduli spaces
nonperturbatively.

\subsection{Mirror symmetry at string tree-level}

Mirror symmetry between the prepotentials of complexified K\"ahler/complex
structure deformations of mirror CY's is naturally formulated in terms of
two specific kinds of coordinates. On the complexified K\"ahler moduli
space of a CY $Y$, the natural coordinates arise through the expansion
around the large volume limit of $Y$. From the perspective of the $N=2$
non-linear sigma model with target space $Y$, they appear in the
exponentials of the three-point correlation functions of chiral primary
operators ${\mathcal O}_a$ associated with classes $A_a$ in $H^{1,1}(Y,
{\mathbb C})$ as computed from the topological A-model
\be\label{buh}
\langle {\mathcal O}_a {\mathcal O}_b {\mathcal O}_c \rangle = \int_Y
A_a \wedge A_b \wedge A_c + \sum_{k_d} N^{k_d}_{abc}\, \e^{{2\pi\I k_d
z^d}}\ .
\ee
Here $k_a$ runs over homology classes of rational curves in $Y$ and
$N_{abc}^{k_d}$ are integers given by the intersection theory on the
moduli space of rational curves, which can be related to the
genus zero Gopakumar--Vafa invariants of $Y$. The expression \eqref{buh}
is the third derivative of the prepotential, which can then be found by
integration. The coordinates $z^a$ agree with the coordinates provided
by the IIB moduli fields \eqref{kahlercoords}.

The canonical coordinates on the moduli space of complex structures of
the mirror CY $X$ are the so-called ``flat'' coordinates arising from
an expansion of the prepotential around the large complex structure
limit of $X$. The coordinates mirror to the $z^a_{\rm IIB}$ appearing
in the large volume expansion of the prepotential \eqref{buh} can be
constructed as follows \cite{Candelas:1990rm,Aspinwall:1993nu}. Using the
so-called monodromy weight filtration \cite{morrison}, one chooses a
symplectic basis  $\{\gamma_{\rm (3)}^\Lambda,\gamma^{\rm (3)}_\Lambda\}$
of $H_3(X,\mathbb Z)$ such that the cycle $\gamma^1_{(3)}$ spans $H_{0,3}
(X,{\mathbb Z})$ (the dual of the class $H^{3,0}(X)$ given by $\Omega(X)$).
It is singled out by having trivial monodromy at the large complex
structure limit. The other cycles $\gamma^a_{(3)}$ constitute a basis of
$H_{1,2}(X,{\mathbb Z})$ and can be shifted in integer multiples of
$\gamma^1_{(3)}$ by monodromy transformations of $H_3(X,{\mathbb Z})$
around the large complex structure limit.\footnote{See \cite{morrison}
for more mathematical details.} The flat (inhomogeneous) coordinates
on the moduli space of complex structures are then the normalized periods
of $\Omega$ introduced in \eqref{specialcoords}. The period integrals
over the dual $B$-cycles $\gamma_\Lambda^{(3)}$ define first derivatives
of the prepotential on the complex structure moduli space of $X$
\be\label{uhh}
F_\Lambda (X) = \int_{\gamma_\Lambda^{(3)}}\! \Omega\ .
\ee
The prepotential is obtained by using homogeneity, $X^\Lambda F_\Lambda
=2 F(X)$. 

In terms of the specific coordinates \eqref{kahlercoords} and
\eqref{specialcoords}, mirror symmetry between the CY's $Y$ and $X$ is
the statement that, under the identification (called the mirror map
\cite{Aspinwall:1993nu})
\be \label{mmap}
z^a_{\rm IIB}\equiv\int_{\gamma^a_{\rm (2)}}\! \big( B_{\rm NS} + \I J
\big) = \frac{\int_{\gamma^a_{(3)}}\,\Omega} {\int_{\gamma^1_{(3)}}\,
\Omega} \equiv z^a_{\rm IIA}\ ,
\ee
the prepotentials derived from \eqref{buh} and \eqref{uhh} agree. The
exponential worldsheet instanton corrections to the former arise from
the expansion of the dual period integrals around the large complex
structure limit.

Furthermore, agreement between the hypermultiplet moduli spaces requires 
the identification of the type IIA and IIB dilatons:
\be\label{dilmirror}
\phi_{\rm IIA} = \phi_{\rm IIB} \ .
\ee
Since mirror symmetry is supposed to work order by order in string
perturbation theory, we will assume this identification to hold
throughout. To simplify our notation we will denote $\tau_2 =
\e^{-\phi_{\rm IIA}} = \e^{-\phi_{\rm IIB}}$ resorting to
\eqref{dilmirror} implicitly.

\subsection{Inclusion of D-branes}

Quantum mirror symmetry between type IIA on $X$ and type IIB on $Y$ also
requires matching the nonperturbative string corrections in the respective
hypermultiplet sectors. The corrections of interest in this paper are the
ones found in \cite{Robles-Llana:2006is} and come from D($-1$) and
D1-brane instantons in type IIB on $Y$. In this subsection we discuss
generic aspects of the mirror symmetry between these objects and D2-branes
in type IIA on $X$ and review the mirror map between the relevant
supersymmetric cycles and RR fields, which we implement at the level of
the LEEA in the next section.

Let us start by considering the D-brane worldvolume actions in type IIB
and type IIA respectively. The D($-1$) instanton couples to the RR scalar
$C_0=\tau_1$, and its action is given by\footnote{Here and henceforth we
are setting $4\pi^2\alpha'=1$. Furthermore, the inverse string coupling
constant is given by $g_s^{-1}=\tau_2$.}
\be\label{nice!!}
S_{\rm D(-1)}=2\pi \tau_2+2\pi\I\, \tau_1\ .
\ee
For D1-brane instantons wrapping a two-cycle with homology $k_a
\gamma^a_{(2)}$, supersymmetry imposes that the map embedding the
worldvolume $\Sigma$ into the two-cycle be holomorphic and that the
field strength $F={\rm d}A$ of the worldvolume gauge field $A$ is such
that $F/2\pi$ is an integral harmonic form on any holomorphic two-cycle
\cite{Marino:1999af,Becker:1995kb}. The worldvolume action $S_{\rm DBI}
+S_{\rm Top}$ for such a D1-instanton is given by \cite{Marino:1999af}
\begin{align}\label{ccoo}
S_{{\rm D1}/k_a\gamma^a_{(2)}} & = 2\pi \int_{\Sigma}\, \tau_2 \,
\sqrt{{\rm det}\big(\hat G + (F/2\pi-\hat B_{\rm NS})\big)}+
2\pi\I\int_\Sigma\,\e^{F/2\pi-\hat{B}_{\rm NS}} \wedge (\hat C_0+
\hat C_2) \notag \\[1.2ex]
& = 2\pi\tau_2\, \sqrt{(k_a t^a)^2+(k_a b^a+n)^2} +2\pi\I \big(k_a c^a -
\tau_1(k_a b^a+n)\big)\ .
\end{align}
In the above formula, $\hat G$ and $\hat B$ denote the pullback of the
space-time fields $G$ and $B$, and $n$ is an arbitrary integer which
encodes the flux of $F$ on the two-cycle (i.e., the first Chern class of
the U(1) bundle). Because of the way the integer $n$ appears in the
topological Chern--Simons coupling in \eqref{ccoo}, it is naturally
associated with D($-1$) charge inside the D1-instanton.

We now turn to the type IIA compactification on $X$. Recall that in the
large volume limit, mirror symmetry aligns the integer homologies
$\bigoplus H_{2k}(Y,{\mathbb Z})$ and $H_3(X,{\mathbb Z})$. More precisely,
the zero homology $H_0(Y,{\mathbb Z})$ is mapped to $H_{0,3}(X,\mathbb Z)$
\cite{Morrison:1995yi}. In the basis \eqref{specialcoords} the homology
$H_{0,3}(X,\mathbb Z)$ is spanned by the three-cycle $\gamma_{(3)}^1$. As
a consequence, a D($-1$)-instanton wrapping a zero-cycle in $Y$ maps to a
D2-instanton wrapping the three-cycle $\gamma_{(3)}^1$ in $X$
\cite{Ooguri:1996ck}. The SYZ construction of mirror symmetry \cite{SYZ}
then allows to identify $\gamma_{(3)}^1$ as a special Lagrangian cycle
with $T^3$ topology. This implies that also the moduli spaces of the
D($-1$) and D2$/\gamma^1_{(3)}$ instantons agree.

For the two-cycles mirror symmetry aligns $H_{1,1}(Y,{\mathbb Z})$ with
$H_{1,2}(X,{\mathbb Z})$ and takes integral bases $\gamma^a_{(2)}$ to
$\gamma^a_{(3)}$ after a suitable linear transformation. As a consequence
of the mixing of zero-forms and two-forms in the coupling of the D1-branes
to $(F/2\pi-\hat B_{\rm NS})$ in \eqref{ccoo}, the mirror of a D1-instanton
wrapping the holomorphic two-cycle in homology $k_a\gamma^a_{(2)}$ with
$n$ units of flux and action \eqref{ccoo} is given by a D2-instanton
wrapping a special Lagrangian cycle in class $k_a\gamma^a_{(3)}+n
\gamma^1_{(3)}$ \cite{Ooguri:1996ck}. Equivalently, a shift of the
$B$-field $k_a b^a\mapsto k_a b^a +n$ is mirror to a monodromy transformation
$k_a \gamma^a_{(3)}\mapsto k_a \gamma^a_{(3)}+n\gamma^1_{(3)}$
\cite{morrison}. 

Based on these considerations the worldvolume action of the D2-brane
instanton mirror to \eqref{nice!!} is given by
\be\label{nicebis}
S_{\rm D2/\gamma^1_{(3)}}= 2\pi \tau_2+2\pi\I\,A^1\ ,
\ee
while the worldvolume action of the D2-brane instanton mirror to
\eqref{ccoo} is
\begin{align}\label{casi}
S_{{\rm D2}/k_a\gamma^a_{(3)}+n\gamma^1_{(3)}} & = 2\pi\tau_2 \Big|\,
\int_{k_a\gamma^a_{(3)}+n\gamma^1_{(3)}}\!\!\Omega\, \Big| - 2\pi\I \int_{k_a
\gamma^a_{(3)} + n \gamma^1_{(3)}}\!\! \hat{C}_3 \notag \\[1.2ex]
& = 2\pi\tau_2 |n+k_a z^a_{\rm IIA}| - 2\pi\I\, \big( k_a
A^a+n A^1\big)\ .
\end{align}
Here we normalized $\Omega$ such that $X^1=1$, $X^a=z^a$, and the fact that
supersymmetry on the worldvolume of the D2-branes constrains $(F/2\pi-
\hat B_{\rm NS})$ to be zero \cite{Marino:1999af}.

Comparing the topological couplings in \eqref{nice!!}, \eqref{nicebis}
and \eqref{ccoo}, \eqref{casi} identifies the remaining pieces of the
mirror map
\be\label{RRm1}
 A^1=\tau_1\ ,\qquad A^a = - (c^a - \tau_1\, b^a)\ .
 \ee
We remark that the shift of the axion by an element of $H^0(Y,{\mathbb Z})$
(i.e., $\tau_1\mapsto \tau_1+1$, which is part of the SL(2,${\mathbb Z}$)
symmetry of the type IIB string on $Y$) gets identified with the shift of
$C_{3}$ by an element of $H^{3,0}(X,{\mathbb Z})$ so that these integral
structures of $Y$ and $X$ match \cite{Morrison:1995yi}.  

We end this section by summarizing the mirror map that relates the
type IIA and type IIB variables
\begin{equation}\label{MM}
\phi_{{\rm IIA}}=\phi_{{\rm IIB}}\ ,\quad A^1 = \tau_1\ ,\quad A^a =
-(c^a - \tau_1 b^a)\ ,\quad   z^a_{{\rm IIA}} = z^a_{{\rm IIB}}\ ,
\end{equation}
where it is understood that type IIB is compactified on the mirror
CY such that the Hodge numbers are interchanged. This is precisely the
closed string mirror map found in \cite{BGHL} and in the conifold limit
\cite{SV}. 

\section{Membrane instantons from mirror symmetry}
\label{sec:5}

In this section, we finally determine the membrane instanton corrections
to the LEEA using the mirror symmetry results
reviewed in the previous section. As explained in Section 2, the LEEA
is completely determined by a single function ${\cal L}$, the
superspace Lagrangian density, or equivalently by the tensor potential
$\chi$, as defined in \eqref{TP}.

\subsection{Resummation of $(p,q)$-string contributions in IIB}
\label{sec:4.1}

The full $(p,q)$-string corrected tensor potential $\chi^{\rm IIB}$ for
type IIB strings compactified on a generic CY $Y$ was found in
\cite{Robles-Llana:2006is} as the $\SL(2,{\mathbb Z})$-invariant
completion of the string tree-level tensor potential $\chi^{\rm
IIB}_\text{tree}$. The latter arises through the c-map by evaluating the
general formula (we use that $F_\Lambda(z)=\partial F/\partial
X^\Lambda (z)$ in the coordinates \eqref{specialcoords})
\be\label{tenspot}
\chi = 4 \,r^0 \tau_2^2\,\mathrm{Im} \left( F_1(z) + \zb^a F_a(z)
\right)
\ee
on the vector prepotential of the T-dual IIA compactification on $Y$
(i.e., the prepotential on the moduli space of complexified K\"ahler
deformations of $Y$) \footnote{Note that the scheme-dependent terms in the prepotential (quadratic in $X^\Lambda$ with real coefficients) do not contribute to the tensor potential. This is analogous to the K\"ahler potential in the vector multiplet sector, where such terms drop out as well.}
\be\label{fkahler}
F_\text{K\"ahler}=F_\text{cl}+F_\text{ws-pert}+F_\text{ws-inst}\ .
\ee
The three terms on the right-hand side of the above equation correspond
respectively to the classical large volume limit prepotential, the
four-loop $\sigma$-model correction, and the contribution from
worldsheet instantons. Substitution of \eqref{fkahler} in
\eqref{tenspot} gives rise to a tensor potential which correspondingly
can be split into three parts. Each of them is then separately completed
into an $\SL(2,{\mathbb Z})$ invariant, resulting in the $(p,q)$-string
corrected tensor potential
\be
\chi^{\rm IIB} = \chi^{\rm IIB}_{\rm cl} + \chi^{\rm
IIB}_{(-1)} + \chi^{\rm IIB}_{(1)}\ ,
\ee
from which the couplings in the effective Lagrangian can be
determined \cite{deWit:2006gn}. Normalized as in \cite{SV}, the
above terms read
\begin{align}\label{IIBTP}
\chi^{\rm IIB}_\text{cl} & = 4\, r^0 \, \tau_2^2 \, \frac{1}{3!}
\,
\kappa_{abc} \, t^a \, t^b \, t^c\ , \notag \\
\chi^{\rm IIB}_{(-1)} & = \frac{r^0 \tau_2^{1/2}}{2 (2 \pi)^3}\,
\chi_E(Y) \, \sideset{}{'}\sum_{m,n}\, \frac{\tau_2^{3/2}}{|m\tau +
n|^3}\ ,
\notag \\
\chi^{\rm IIB}_{(1)} & = - \frac{r^0 \tau_2^{1/2}}{(2 \pi)^3}\,
\sum_{ k_a } n_{k_a} \sideset{}{'} \sum_{m,n} \frac{\tau_2^{3/2}}
{|m\tau + n|^3}\, \big( 1 + 2 \pi |m\tau + n|\, k_a t^a \big) \,
\e^{-S_{m,n}}\ .
\end{align}
The above expressions are written in terms of the fields introduced in
subsection 3.1. $\chi_E(Y)$ and $\kappa_{abc}$ are the Euler number and
classical triple intersection form on $Y$ respectively, and the primes in
the sums indicate that the $(m,n)=(0,0)$ term is excluded. Moreover
\be\label{Smn}
  S_{m,n} = 2\pi k_a \big( |m\tau + n|\, t^a - \I m\, c^a - \I n\, b^a
  \big)\ ,
\ee
is the action of a $(p,q)$-string wrapped gcd$(m,n)$ times around
a holomorphic 2-cycle in homology ${k_a}$ with respect to the
basis $\gamma^a_{(2)}$ introduced in Section 3. 

The different terms in \eqref{IIBTP} have then the following
interpretation: $\chi_{\rm cl}$ is the classical contribution
arising through the c-map from the large volume limit
$F_\text{cl}$ of $F_\text{K\"ahler}$ in \eqref{fkahler}. It is by
itself modular invariant, reflecting the $\SL(2,{\mathbb R})$
invariance of the classical action. $\chi_{(-1)}$, the
$\SL(2,{\mathbb Z})$ completion of the perturbative $\sigma$-model
correction, encodes the corrections to the couplings in the LEEA
arising from $(p,q)$-string maps in which the $(p,q)$-string
worldvolume is taken to a point in the CY\@. It includes the
one-loop string corrections, together with an infinite series of
D($-1$) instanton corrections. Finally $\chi_{(1)}$, the
$\SL(2,{\mathbb Z})$ completion of the worldsheet instanton
contributions, encompasses the instantons of the $(p,q)$-string
for which the latter's worldvolume wraps holomorphically embedded
two-cycles $k_a\gamma^a_{(2)}$ in the CY (note that instantons of
the $(1,0)$-string are the worldsheet instantons). The
multiplicity of these cycles in each homology class is
characterized by the Gopakumar--Vafa integers $n_{k_a}$.

In order to map the above expressions to the mirror type IIA
compactification, we first perform a Poisson resummation to go to
an equivalent representation, which has the virtue of making the
role of D($-1$)- and D1-brane instantons (i.e., their standard
topological couplings to the RR fields) manifest. 

As shown in \cite{Robles-Llana:2006is} one can resum $\chi_{(-1)}^{\rm IIB}$ into
\be\label{E32}
  \chi^{\rm IIB}_{(-1)}   = \frac{\chi_E(Y)}{2(2\pi)^3}\, r^0 \sqrt{\tau_2}\,\Big[\, 2
  \zeta(3)\, \tau_2^{3/2} + \frac{2\pi^2}{3}\, \tau_2^{-1/2} + 8\pi\,
  \tau_2^{1/2}\!\!\!\! \sum_{m\neq 0,n>0} \Big| \frac{m}{n} \Big|\,
  \e^{2\pi \I mn \tau_1}\, K_1(2\pi |mn| \tau_2)\Big]\, .
\ee
The first term arises from worldsheet perturbation theory, the second from the one-loop string correction, while after expanding the Bessel function at large $\tau_2$ (weak
coupling) one observes that the sum is the contribution from
the D($-1$) instantons to the tensor potential, including the
perturbative corrections around the instanton background:
\begin{align}
\chi^{\rm IIB}_{\rm D(-1)} = \frac{r^0
\tau_2}{2\pi^{3/2}}\,\chi_E(Y)\!\! \sum_{m\neq 0,n>0} &
\left|\frac{m}{n}\right|\,\left[\sum_{k=0}^\infty\,
\frac{\Gamma(3/2+k)}{k!\,
\Gamma(3/2-k)}\, (4\pi|mn|\tau_2)^{-k-1/2}\right]\,\e^{-2\pi|mn|\tau_2 + 2\pi \I mn
\tau_1}\ .
\end{align}

In a similar spirit we split off the worldsheet instanton
contribution $(m=0)$ in $\chi_{(1)}$:
\be \chi^{\rm IIB}_{(1)} =  \chi^{\rm IIB}_\text{ws-inst}
+\chi^{\rm IIB}_\text{D1-inst}\ .\ee
We again perform a Poisson resummation on the unrestricted integer
$n$ in $\chi_\text{D1-inst}$ (see Appendix A) and obtain
\be\label{D1res} \chi^{\rm IIB}_\text{D1-inst} = - \, \frac{r^0
\tau_2}{2 \pi^2} \sum_{k_a} n_{k_a}\!\! \sum_{m \not = 0, n \in
\Zom} \frac{|z+n|}{|m|}\, K_1(2 \pi |m \tau_2| |z+n|) \, \e^{2 \pi
\I m (c-\tau_1 (b + n))}\ , \ee
where we have denoted $z=k_a z^a$, $b=k_a b^a$ and $c=k_a c^a$.
Expanding the Bessel function for large $\tau_2$, we obtain
\begin{align}
\chi^{\rm IIB}_\text{D1-inst}= -\frac{r^0 \tau_2}{4\pi^2}\,
\sum_{k_a}n_{k_a}\!\! \sum_{m\neq 0,n\in{\mathbb Z}}\, &
\frac{|z+n|^{1/2}}{|m|^{3/2}}\, \Bigg[ 1 + \sum_{k=1}^{\infty}\,
\frac{\Gamma(3/2+k)}{k!\,
\Gamma(3/2-k)}\, (4\pi |m\tau_2||z+n|)^{-k} \Bigg] \times \notag \\
&\times \mathrm{exp} \big[ - 2\pi |m\tau_2||z+n|+ 2\pi\I m (c-\tau_1
(b+n)) \big]\ .
\end{align}
The argument of the exponential reproduces the worldvolume
action expected from \eqref{ccoo} for  D1-instantons (with $n$ units
of flux) wrapped $m$ times around $k_a\gamma^a_{(2)}$.

\subsection{Mirror symmetry and membrane instantons in type IIA}

Using the mirror map discussed in Section \ref{sec:mirror}, the
expressions obtained in the previous subsection for type IIB strings
compactified on $Y$ can now be mapped to a type IIA
compactification on the mirror CY $X$.

We first consider the string tree-level tensor potential
$\chi^{\rm IIB}_\text{tree}$ for type IIB compactified on $Y$. It
consists of three parts \be\label{chitree} \chi_\text{tree}^{\rm
IIB}(z^a_{\rm IIB})=\chi^{\rm IIB}_\text{cl} +\chi^{\rm
IIB}_\text{ws-pert}+\chi^{\rm IIB}_\text{ws-inst} \ee arising from
the three terms in the prepotential $F_\text{K\"ahler} (z^a_{\rm
IIB})$ given in \eqref{fkahler}. In Section 3 we saw that, under
the identification \eqref{mmap}, this prepotential is equal to the
prepotential on the moduli space of complex structure deformations
of $X$. We then conclude that \eqref{chitree} equals the type
IIA tree-level tensor potential \eqref{tenspot} as computed from
the prepotential $F_{\rm cs}(z^a_{\rm IIA})$ on the moduli space of
complex structures on the mirror CY $X$. Furthermore, the mirror
map in eq.\ \eqref{mmap} provides the map between the NS-NS fields
in the type IIB and IIA compactifications.

Next, we turn to matching the string one-loop term
$\chi_\text{loop}$. In \cite{RSV} it was shown that this
correction comes with opposite signs in IIA and IIB
compactifications on the same CY (consistent with the fact that
the tensor structures involved in the dimensional reduction of the
ten dimensional $R^4$ terms come with different signs in the type
IIA and IIB theories \cite{Antoniadis:1997eg}). Taking into account that
$\chi_E(Y)=-\chi_E(X)$ we conclude that this term is left
invariant under mirror symmetry
\be \chi^{\rm IIA}_\text{loop}=-\frac{1}{24\pi}\, r^0 \,\chi_E(X)=
\frac{1}{24\pi} \, r^0 \, \chi_E (Y)=\chi^{\rm IIB}_\text{loop}\ .
\ee

Finally, using the discussion in Section \ref{sec:mirror}, the
nonperturbative contributions $\chi^{\rm IIB}_\text{D($-1$)}$ and
$\chi^{\rm IIB}_\text{D1}$ can also easily be mapped to the type
IIA compactification.

The D($-1$) instantons, whose contribution is encoded in
$\chi^{\rm IIB}_\text{D($-1$)}$ are mapped to D2-brane instantons
wrapping the cycle $\gamma^1_{(3)}$ with associated tensor potential
\be \label{Dm1res}  \chi^{\rm IIA}_{{\rm D2}/\gamma^{1}_{(3)}} = -
\, \frac{r^0 \tau_2}{2 \pi^2} \left(\frac{\chi_E(X)}{2} \right)\!
\sum_{m,n\neq 0} \left| \frac{n}{m} \right| \e^{- 2 \pi \I mn A^1}
\, K_1(2 \pi |mn| \tau_2)\ , \ee
which follows directly after substitution of the mirror map \eqref{MM} in
$\chi^{\rm IIB}_{\rm D(-1)}$.

Similarly, D1 instantons wrapping holomorphic cycles in homology
$k_a \gamma^a_{(2)}$ map to D2-brane instantons wrapping special
Lagrangian cycles in homology $k_a\gamma^a_{(3)}+n\gamma^1_{(3)}$, whose
contribution to the tensor potential is again found by using the
mirror map in $\chi_{\rm D1}^{\rm IIB}$ given by \eqref{D1res}. This
results in
\be\label{buff}
\chi^{\rm IIA}_{{\rm D2}/\gamma^a_{(3)}+n\gamma_{(3)}^1} = - \frac{r^0
\tau_2}{2\pi^2}\, \sum_{k_a}\, n_{k_a}\! \sum_{m \not = 0, n \in \Zom}
\frac{|z+n|}{|m|}\, K_1(2 \pi |m \tau_2| |z+n|)\, \e^{- 2 \pi \I m k_a
A^a}\, \e^{-2 \pi \I m n A^1}\ ,
\ee
where $z=k_a z^a$ as above.

We can now combine the potentials \eqref{Dm1res} and \eqref{buff}
by introducing vectors
\be k_{\Lambda} = \big( n, k_a\big)\ , \qquad z^\Lambda = \big( 1
, z^a \big)\ , \qquad A^\Lambda = \big( A^1 , A^a \big)\ , \ee
$\Lambda\in\{1, a\}$, and write the full nonperturbative IIA
result as a sum over vectors $k_\Lambda$ weighted by instanton
numbers $n_{k_\Lambda}$ as follows:
\be\label{D2instres}
\chi^{\rm IIA}_{{\rm A-D2}} = - \frac{r^0 \tau_2}
{2\pi^2}\, \sum_{k_\Lambda} n_{k_\Lambda} \sum_{m \not = 0}
\frac{|k_\Lambda z^\Lambda|}{|m|} \,  K_1\big( 2 \pi \tau_2 \, |m
\, k_\Lambda z^\Lambda| \big) \, \e^{-2 \pi \I m k_\Lambda
A^\Lambda}\ .
\ee
Here, the sum over $k_a$ now includes the zero-vector $k_a=0$, but
$k_\Lambda =0$ is excluded. The type IIA instanton numbers read
\be\label{electvec} n_{(n,k_a=0)} = \frac{1}{2}\chi_E(X)\ ,\qquad n_{(n,k_a)}
=n_{k_a}\ \text{as in type IIB} \, . \ee
This formula captures all type IIA membrane instanton contributions
arising from Euclidean D2-branes wrapping the $A$-cycles of the CY.

Collecting our results, we write our final expression for the
complete tensor potential encoding the $A$-cycle membrane
instanton corrected LEEA for type IIA compactifications on a
generic CY $X$ as
\begin{align}
\chi^{\rm IIA} & = \chi_\text{tree}^{\rm IIA} + \chi^{\rm
IIA}_\text{loop} + \chi^{\rm IIA}_{{\rm A-D2}} \notag \\*[2pt] & =
4\, r^0 \tau_2^2\, \mathrm{Im} \left[ F_1(z_{\rm IIA}) + \zb_{\rm
IIA}^a F_a (z_{\rm IIA})\right] - \frac{1}{24\pi}\, r^0\,
\chi_E(X) \notag \\* &~~~ - \, \frac{r^0 \tau_2}{2 \pi^2}\,
\sum_{k_\Lambda} n_{k_\Lambda} \sum_{m \not = 0} \frac{|k_\Lambda
z^\Lambda|}{|m|}\, K_1\big( 2 \pi \tau_2 \, |m\, k_\Lambda
z^\Lambda| \big)\, \e^{-2 \pi \I m k_\Lambda A^\Lambda}\ .
\end{align}
The latter formula implies that D2-brane instanton corrections to the
LEEA for type IIA compactified on a CY $X$ are determined by two kinds
of topological invariants: the Euler characteristic $\chi_E(X)$ of the
CY $X$, and the Gopakumar--Vafa invariants $n_{k_a}$ of the mirror CY
$Y$.

\subsection{Superspace description of the instanton corrections}
%

As already discussed in Section \ref{sec:superspace}, the hypermultiplet
sector arising form a type II string compactification on a generic CY
can be encoded by the tensor potential $\chi$ or the superspace density
$\cL$. When implementing symmetries like the $\SL(2,\Zom)$ invariance of
the type IIB string, it is natural to work with $\chi$ since a symmetry of
the effective action directly translates into an invariance of $\chi$.
When trying to generalize the results above along the lines of the
conjecture made in \cite{Anguelova:2004sj} to include also NS5-brane
instantons, or motivated by making contact with the hybrid formalism on
the heterotic string side, it is desireable to also have a description of
the instanton corrections in terms of the projective superspace density
$\cL$. Since the corresponding derivation is somewhat technical and not
very illuminating, we restrict ourselves to giving the final result,
while the details are collected in Appendix \ref{App:B}. 

Defining the vector
\be
\tilde{\eta}^\Lambda(\zeta) = \big( n \eta^0 + m \eta^1 \, , \,
\eta^a \big)\ , 
\ee
eqs.\ \eqref{Lcont2} and \eqref{cont1} can conveniently be combined into 
\be\label{eq:Dcont2}
 \cL = {\rm Im} \, \bigg[ \frac{\I}{4 \pi^3} \, \sum_{k_\Lambda}
n_{k_\Lambda}\! \sum_{m > 0,n \in \Zom} \,   \oint_{\cC_{m,n}} \frac{\rmd
\zeta}{2 \pi \I \zeta} \, \frac{1}{m^2} \, \frac{(\eta^0)^2}{n
\eta^0 + m \eta^1}\, \e^{-2 \pi \I m k_\Lambda \tilde{\eta}^\Lambda /
\eta^0} \bigg]\ . 
\ee
Here, the contours $\cC_{m,n}$ enclose a zero $\hat{\zeta}_+$
of $\zeta (n \eta^0 + m \eta^1)$, and $n_{k_\Lambda}$ is given
in \eqref{electvec}. Evaluating the contour integral using
Cauchy's Integral formula and computing $\chi$ via \eqref{TP},
\eqref{eq:Dcont2} reproduces $\chi^{\rm IIA}_\text{A-D2}$ given 
in \eqref{D2instres}. Note that since both D($-1$) and D1 instanton
corrections in IIB and $A$-type membrane corrections are encoded in the
same superspace Lagrangian, our results provide the superspace
description for both types of corrections simultaneously.

\section{Discussion and conclusions}

In this paper we have used mirror symmetry to determine and sum up
a class of membrane instanton corrections, drawing on previous results on
D($-1$) and D1 instanton corrections for type IIB strings compactified on
a generic Calabi--Yau threefold \cite{Robles-Llana:2006is}. These
nonperturbative corrections encompass the instanton corrections of
Euclidean D2-branes wrapping three-cycles ($A$-cycles) dual to the
supersymmetric zero- and two-cycles of the mirror Calabi--Yau and can
conveniently be encoded either in terms of the tensor potenial
\eqref{D2instres} or the projective superspace density \eqref{eq:Dcont2}.
The resulting formulas are completely fixed by the Euler number and the
genus zero Gopakumar--Vafa invariants of the mirror Calabi--Yau. On the
type IIA side, the instanton corrections thereby contain contributions
from both rigid ($b_1=0$) and non-rigid ($b_1\not=0$) three-cycles, since,
according to the SYZ interpretation of mirror symmetry \cite{SYZ}, the
cycle mirror to the zero-cycle has first Betti number $b_1=3$. Thus, the
mirror result automatically comprises the integrations over the instanton
moduli space including an appropriate measure.

The results obtained in this paper rely on the dual corrections on the
type IIB side and nonperturbative mirror symmetry. It would be interesting
to rederive these corrections from a microscopic IIA computation. This
will require an interpretation of the Gopakumar--Vafa invariants of the
mirror Calabi--Yau in terms of properties of special Lagrangian
three-cycles. A proposal for a topological quantity capturing these
properties has recently been made by Joyce \cite{Joyce}, and it would be
interesting to establish a connection between this proposal and mirror
symmetry results.

A natural question is how to generalize our results so as to include
the remaining membrane instanton corrections arising from the Euclidean
D2-branes wrapping supersymmetric $B$-cycles and possibly supersymmetric
combinations involving both $A$- and $B$-cycles. Fig.\ \ref{zwei}
suggests that their form can be deduced from imposing electric-magnetic
duality invariance, i.e., invariance under discrete basis transformations
in $H_3(X,\Zom)$, on the LEEA. From the structure of \eqref{D2instres}
it is clear that, besides the Gopakumar--Vafa invariants and the Euler
number, the full result will also involve additional topological
invariants of the Calabi--Yau which encode a suitable multiplicity of
the $B$-cycles (or their dual four- and six-cycles on the mirror
Calabi--Yau). Furthermore, implementing these corrections in the
low-energy effective action will require a generalization of the tensor
multiplet framework employed in this paper, since the corresponding
corrections will break (some of) the remaining shift symmetries which
are manifest when working with tensor multiplets. Thus, finding the
complete membrane instanton corrected low-energy effective action will
require progress in formulating off-shell $N=2$ supergravity actions as well as
a better understanding of the topological invariants on the
compactification manifolds. We hope to return to these questions in
the future.

\vspace{10mm} \noindent {\bf Acknowledgements \\} 
We thank Martin Ro\v{c}ek for collaboration during the initial stages of
this project at the fourth Simons Workshop in Mathematics and Physics.
Furthermore, we thank Frederik Denef, Thomas Grimm, Jean Dominique
L\"ange and Pierre Vanhove for useful discussions. DRL is supported by
the European Union RTN network MRTN-CT-2004-005104. FS is supported by
the European Commission Marie Curie Fellowship no.\ MEIF-CT-2005-023966. 
SV thanks the Galilei Galileo Institute for Theoretical Physics for its
hospitality and partial support. Further support is acknowledged from
INTAS contract 03-51-6346.

%
\begin{appendix}
\section{Poisson resummation revisited}
\label{App:A.1}
%
In this appendix we give the identities required for resumming the D1
instanton corrections in Subsection \ref{sec:4.1}, following closely and
extending Appendix B of \cite{SV}. They are based on the Poisson
resummation formula
\be\label{PRF}
\sum_{n \in \Zom} f(x+na) = \frac{1}{a} \sum_{n \in \Zom} \tilde{f}(2
\pi n /a) \, \e^{2 \pi \I n x/a} \quad \text{with} \quad \tilde f(k) =
\int_{-\infty}^{\infty}\! {\rm d}x\, f(x)\, \e^{-ikx}\ .
\ee

In order to establish eq.\ \eqref{D1res}, we apply this identity to
\be\label{eq:D1-inst} \chi^{\rm IIB}_{\rm D1-inst} = - \frac{r^0
\tau_2^{2}}{(2 \pi)^3}\, \sum_{ k_a } n_{k_a}  \sum_{m \not = 0}
\, \Bigg[ \, \sum_{n \in \Zom} \frac{1}{|m\tau + n|^3}\, \big( 1 +
2 \pi |m\tau + n|\, k_a t^a \big)\, \e^{-S_{m,n}} \Bigg] \ee
with $S_{m,n}$ given in \eqref{Smn}. This requires performing an
inverse Fourier transform of the summand appearing in square brackets.
Comparing to the general formula \eqref{PRF}, we identify
\be\label{A.11} \tilde{f}(2 \pi n) = \frac{(2 \pi)^3}{\big(
\alpha^2 + (2 \pi n + \gamma)^2 \big)^{3/2}}\, \big( 1 + \sqrt{
\alpha^2 + (2 \pi n + \gamma)^2 }\, t \big)\, \e^{- \sqrt{ \alpha^2
+ (2 \pi n + \gamma)^2 } \, t} \, \e^{2 \pi \I m c}\ , \ee
with $a = 1$, $x = k_a b^a$, and we set $\alpha = 2 \pi m \tau_2$,
$\gamma = 2 \pi m \tau_1$, $t = k_a t^a$, $c = k_a c^a$.

The inverse Fourier transform of this expression can be found making
the following observation. In \cite{SV} we gave the following formula for
Fourier Cosine transformations \cite{IT}
\be\label{FCT1}
\int_0^\infty\! {\rm d}x\ \frac{1}{\sqrt{x^2 + \alpha^2}}\, \e^{-\beta\,
\sqrt{x^2 + \alpha^2}} \cos(xy) = K_0 \big( \alpha\, \sqrt{\beta^2 +
y^2}\, \big)\ .
\ee
Taking a derivative with respect to the parameter $\alpha$ leads to the
identity
\be\label{FCT2}
\int_0^\infty\! {\rm d}x\ \frac{1 + \beta \, \sqrt{x^2 +
\alpha^2} }{\sqrt{x^2 + \alpha^2}^3} \, \, {\rm e}^{-\beta
\,\sqrt{x^2 + \alpha^2}} \cos(xy) = \, \frac{1}{\alpha} \,
\sqrt{\beta^2 + y^2}\ K_1 \big( \alpha \sqrt{\beta^2 + y^2}\,
\big)\ .
\ee
Comparing the integrand appearing on the LHS to \eqref{A.11}, we
observe that this is precisely the Fourier transform required for
resumming the D1 instantons in \eqref{eq:D1-inst}. Carrying out
the resummation then leads to the result \eqref{D1res}.
%
\section{\!Formulating the instanton corrections in projective superspace}
\label{App:B}
%
This appendix contains the derivation of the contour integral
representation of the instanton corrections given in eq.\
\eqref{eq:Dcont2}. Since both D($-1$) and D1 instanton corrections in IIB
and $A$-type membrane corrections are encoded in the same superspace
Lagrangian, our results provide the superspace description for both types
of corrections simultaneously.

The derivation of this contour formulation is complicated by the
fact that eq.\ \eqref{TP} cannot be solved for $\cL$ without
performing a nontrivial integration. We will then employ the
following strategy. Starting from the instanton contributions in
\eqref{IIBTP}, we use the relation \eqref{cLrel} to compute
$\cL_{x^I x^J}$. Using Mathematica, the resulting expressions can
be integrated twice, and one can explicitly check that the
resulting function $\cL$ satisfies the supersymmetry constraint
\eqref{Llap}. The corresponding contour integral representation is
then found by trial and error. We first derive the contour
formulation for the D($-1$) instantons before turning to the D1
instantons in the next subsection.
\subsection{The D(--1) instanton sector}
We start by splitting the D($-1$) instanton contribution
$\chi^{\rm IIB}_{(-1)}$ into the perturbative worldsheet
corrections $m=0,n\not=0$ and the contribution from D($-1$)
instantons
\be \chi =  \frac{r^0}{2 (2 \pi)^3}\,\, \chi_E\!\! \sum_{m \not =
0,n \in \Zom}\, \frac{\tau_2^{2}}{|m\tau + n|^3}\ . \ee
Substituting into \eqref{cLrel}, we find\footnote{Alternatively
this result can be derived by computing $\cL_{x^1 x^1}$ based on
$\chi^{\rm IIB}_{\rm D(-1)}$ given in \eqref{E32} and performing a
Poisson resummation.}
\be\label{Lxx3} \cL_{x^1 x^1} = - \frac{\chi_E}{(2 \pi)^3 r^0}
\sum_{m > 0, n \in \Zom} \, \frac{m^2 \tau_2^2 - 2 \, (m \tau_1 +
n)^2}{\big[m^2 \tau_2^2 + (m \tau_1 + n)^2 \big]^{5/2}}\ . \ee
Note that this result does not contain the perturbative one-loop
correction since, by virtue of the expansion \eqref{E32}, this
does not contribute to $\cL_{x^1 x^1}$. We now read $\tau_1,
\tau_2$ as functions of the tensor multiplet scalars $v, \vb, x$
(cf.\ eq.\ \eqref{vocabIIB}) and integrate $\cL_{x^1 x^1}(x,
v, \vb)$ with respect to $x^1$. In principle this integration
could give rise to two nontrivial integration functions $g_i(v^0,
\vb^0, v^1, \vb^1, x^0)$ multiplied by terms independent and
linear in $x^1$. These integration functions encode the one-loop
correction which arises from the known contour integal expression
\eqref{oneloopL} and will be set to zero in the following. The
result can then be written in the following, suggestive form
\be\label{LD-1} \cL =   \frac{\chi_E(Y)}{2 (2 \pi)^3}  \sum_{m >
0, n \in \Zom}\, \frac{1}{m^2} \, \frac{1}{|n \vr{0} + m \vr{1}|}
\, \left( \eta^0(\hat{\zeta}_+)^2 + \eta^0(\hat{\zeta}_-)^2
\right)\ . \ee
Here
\begin{align}
\eta^I(\hat{\zeta}_\pm) =  x^I - \frac{n x^0 + m x^1}{2} & \left[
\frac{v^I}{n v^0 + m v^1} + \frac{\vb^I}{n \vb^0 + m \vb^1}
\right] \notag \\[2pt]
\mp\, \frac{|n \vr{0} + m \vr{1}|}{2} & \left[ \frac{v^I}{n v^0 + m
v^1} - \frac{\vb^I}{n \vb^0 + m \vb^1} \right] \, ,
\label{zetaeta}
\end{align}
which is valid for all tensor multiplets $I = 0, \ldots ,
h_{1,1}(Y)+1$. One can then check by explicit calculation that
\eqref{LD-1} satisfies the supersymmetry constraint $\cL_{x^I x^J}
+\cL_{v^I \vb^J} = 0$, eq.\ \eqref{Llap}. Thus $\cL$ must have a
representation in terms of a contour integral \eqref{cint}.

Indeed one can verify that \eqref{LD-1} can be obtained from
evaluating
\be\label{Lcont2} \cL = \mathrm{Im} \left[ \frac{\I \chi_E}{(2
\pi)^3}
 \sum_{m > 0,n \in \Zom } \,
 \oint_{\cC_{m,n}} \,
\frac{\rmd \zeta}{2 \pi \I \zeta} \, \frac{1}{m^2} \;
\frac{(\eta^0)^2}{n \eta^0 + m \eta^1} \right]\, , \ee
with the contour $\cC_{m,n}$ enclosing the zero $\hat{\zeta}_+$ of
$\zeta (n \eta^0 + m \eta^1)$,
\be\label{zetap} \hat{\zeta}_{\pm} = \frac{1}{2 (n \vb^0 + m
\vb^1)} \left[ (n x^0 + m x^1) \mp |n \vr{0} + m \vr{1}|\, \right]
\, . \ee

In order to verify this statement explicitly, one uses Cauchy's
integral formula. Noting that the denominator appearing in
\eqref{Lcont2} may be written as
\be\label{eq:denom} \zeta (n \eta^0 + m \eta^1) = - (n \vb^0 + m
\vb^1) \, (\zeta - \hat{\zeta}_+) \, (\zeta - \hat{\zeta}_-)\ ,
\ee
one finds that the integrand has a simple pole at
$\zeta=\hat{\zeta}_+$. Evaluating \eqref{eq:denom} at this pole
gives $\zeta (n \eta^0 + m \eta^1)|_{\zeta = \hat{\zeta}_+} = |n
\vr{0} + m \vr{1}|$, while $\eta^I(\zeta)|_{\zeta =
\hat{\zeta}_+}$ precisely gives rise to \eqref{zetaeta}.

\subsection{Superspace description of D1 instantons}
%
Deriving the contour integral representation for the D1 instantons
completely parallels the previous computation. We start from the
D1 instanton contribution contained in \eqref{IIBTP}
\be\label{D1inst} \chi = - \frac{r^0 \tau_2^{1/2}}{(2 \pi)^3}\,
\sum_{ k_a  } n_{k_a} \sum_{m \not = 0,n \in \Zom}
\frac{\tau_2^{3/2}}{|m\tau + n|^3}\,
  \big( 1 + 2 \pi |m\tau + n|\, k_a t^a \big) \, \e^{-S_{m,n}}\ .
\ee
Using eq.\ \eqref{cLrel} we compute
\be\label{LxxD1} \cL_{x^a x^b} = \frac{1}{2 \pi r^0} \sum_{k_a}
n_{k_a}\, k_a k_b\! \sum_{m \not = 0,n \in \Zom} \frac{1}{|m\tau +
n|}\, \e^{-S_{m,n}}\ . \ee
Again taking the Poincar\'e fields as functions of $v, \vb, x$, we
note that the scalar fields $x^a$ enter into this expression as
linear terms in $S_{m,n}$ only. This allows to integrate $\cL_{x^a
x^b}$ with respect to $x^a$, $x^b$.  In this step we again dropped the integration functions independent of and linear in $x^a$ which encode the one-loop and D(-1) brane instanton corrections already determined. The resulting expression for $\cL$ is
rather cumbersome, but can be written in terms of $\eta^I(\hat{\zeta}_\pm)$  similar as in
\eqref{LD-1}.

Using the techniques of the previous subsection, one can then
explicitly check that the resulting function $\cL$ has the contour
integral description
\be\label{cont1} \cL = {\rm Im} \, \bigg[ \frac{\I}{4 \pi^3} \,
\sum_{k_a} n_{k_a}\! \sum_{m > 0,n \in \Zom} \,
\oint_{\cC_{m,n}} \frac{\rmd \zeta}{2 \pi \I \zeta} \,
\frac{1}{m^2} \, \frac{(\eta^0)^2}{(n \eta^0 + m \eta^1)}\, \e^{-2
\pi \I m k_a \eta^a / \eta^0} \bigg]\, , \ee
where the contours $\cC_{m,n}$ are again taken around the zero
$\hat{\zeta}_+$ of $\zeta (n \eta^0 + m \eta^1)$. Evaluating the
contour integral via Cauchy's integral formula, it is then
straightforward, but tedious, to check that \eqref{cont1}
correctly reproduces \eqref{LxxD1}.

\end{appendix}

\end{document}